\documentclass[final,5p,times,twocolumn,preprint]{elsarticle}
\usepackage{amssymb,amsmath}
\allowdisplaybreaks
\newcommand{\be}{\begin{eqnarray}}
\newcommand{\ee}{\end{eqnarray}}
\newcommand{\beq}{\begin{equation}}
\newcommand{\eeq}{\end{equation}}
\begin{document}
\begin{frontmatter}
\vspace{-1ex}
\title{\begin{flushright}
{\footnotesize JLAB-THY-18-2675}\\
\end{flushright}
Accurate nucleon electromagnetic form factors \\
from dispersively improved chiral effective field theory}
\author{J. M. Alarc\'on}
\author{C. Weiss}
\address{Theory Center, Jefferson Lab, Newport News, VA 23606, USA}
\begin{abstract}
We present a theoretical parametrization of the nucleon electromagnetic form factors (FFs) based
on a combination of chiral effective field theory and dispersion analysis.
The isovector spectral functions on the two-pion cut are computed using elastic unitarity,
chiral pion-nucleon amplitudes, and timelike pion FF data. Higher-mass isovector 
and isoscalar $t$-channel states are described by effective poles, whose strength is 
fixed by sum rules (charges, radii). Excellent agreement with the spacelike proton and neutron 
FF data is achieved up to $Q^2 \sim 1$ GeV$^2$. Our parametrization provides
proper analyticity and theoretical uncertainty estimates and can be used for low-$Q^2$ 
FF studies and proton radius extraction.
\end{abstract}
\end{frontmatter}
\section{Introduction}
The electromagnetic form factors (EM FFs) parametrize the transition matrix element of the 
EM current between nucleon states and represent basic characteristics of nucleon structure.
The FFs at spacelike momentum transfers $Q^2 \lesssim 1$ GeV$^2$ 
have been measured in a series of elastic electron scattering 
experiments \cite{Perdrisat:2006hj,Punjabi:2015bba,Ye:2017gyb}, most recently at 
the Mainz Microtron (MAMI) \cite{Bernauer:2010wm,Bernauer:2013tpr,Mihovilovic:2016rkr}
and at Jefferson Lab \cite{Crawford:2006rz,Paolone:2010qc,Zhan:2011ji}.
The derivative of the proton electric FF at $Q^2 = 0$ (charge radius) is also determined 
with high precision in atomic physics experiments. Discrepancies between results
obtained with different methods have raised interesting questions concerning the 
precise value of the proton charge radius and the $Q^2 \rightarrow 0$ extrapolation 
of the elastic scattering data \cite{Pohl:2010zza,Carlson:2015jba,Krauth:2017ijq}. 
Besides their importance for nucleon structure, the EM FFs are needed as an input in
other areas of study, such as precision measurements of quantities used to test the 
Standard Model. 

The experiments and applications require a theoretical description
of the FFs that covers a broad range $Q^2 \sim$ few GeV$^2$ and controls the behavior 
in the $Q^2 \rightarrow 0$ limit (higher derivatives). This can be accomplished using
the framework of dispersion theory, which incorporates the analytic properties 
of the FF in the momentum transfer. Dispersive parametrizations of the nucleon FFs have 
been constructed using empirical spectral functions, determined by amplitude analysis 
techniques and fits to the FF data \cite{Hohler:1974eq,Hohler:1976ax,Belushkin:2006qa,Lorenz:2012tm}. 
It would be desirable to have a dispersive parametrization that is based on first-principles 
dynamical calculations and permits theoretical uncertainty estimates.

In recent work we developed a method for computing the spectral functions of
nucleon FFs on the two-pion cut using a combination of $\chi$EFT and amplitude analysis
(dispersively improved $\chi$EFT, or DI$\chi$EFT) \cite{Alarcon:2017ivh,Alarcon:2017lhg}. 
The spectral functions are constructed using the elastic unitarity condition. 
The $N/D$ method is used to separate the $\pi\pi$ rescattering effects (contained in 
the pion timelike FF) from the coupling of the $\pi\pi$ system to the nucleon 
(calculable in $\chi$EFT with good convergence). The method permits computation
of the two-pion spectral functions up to masses $\sim$1 GeV$^2$ with controled accuracy. 
In Ref.~\cite{Alarcon:2017lhg} the computed spectral functions in LO, NLO, and partial N2LO,
accuracy were used to study the FFs at low $Q^2$ ($<$0.5 GeV$^2$ for $G_E$, $<$0.2 GeV$^2$ 
for $G_M$) and their derivatives.

In this letter we use DI$\chi$EFT to calculate the nucleon FFs up to $Q^2 \sim 1$ 
GeV$^2$ (and higher) and construct a dispersive parametrization of the FFs with theoretical
uncertainty estimates. This is achieved by extending our previous calculations in two aspects:
(a) We partially include N2LO chiral loop corrections in the isovector magnetic spectral function, 
by parametrizing them in a form similar to the N2LO corrections in the electric case.
This brings the calculation of electric and magnetic isovector FFs 
up to the same order. (b) We account for higher-mass $t$-channel states in the 
spectral functions (isovector and isoscalar) by parametrizing them through effective poles, 
whose strength is determined by sum rules (charges, magnetic moments, radii). This allows
us to extend the dispersion integrals to higher masses and compute the spacelike FFs
up to higher $Q^2$. We obtain an excellent description of $G_E$ and $G_M$
up to $Q^2 \sim 2$ GeV$^2$ with controled theoretical accuracy. Our results
represent genuine theoretical predictions, as no fits are performed and no spacelike FF 
data are used in determining the parameters. In the following we describe the calculation 
and results and discuss potential applications of our FF parametrization.
\section{Method}
The FFs are analytic functions of the invariant momentum transfer $t \equiv -Q^2$ 
and satisfy dispersion relations
\beq
G^{p,n}_i (t) \; = \; \frac{1}{\pi} \int_{t_{\rm thr}}^\infty dt' \; 
\frac{\text{Im} \, G^{p,n}_i(t')}{t' - t - i 0}
\hspace{1cm} (i = E, M) .
\label{Eq:Dispersive_representation}
\eeq
They allow one to reconstruct the spacelike FFs from the spectral functions 
$\text{Im} \, G^{p,n}_i (t')$ on the cut at $t' > t_{\rm thr}$. For theoretical 
analysis one uses the isovector and isoscalar combinations, $G^{V,S}_i \equiv \frac{1}{2} 
(G_i^p \mp G_i^n) \; (i = E, M)$. In the isovector FF the lowest singularity is the
two-pion cut with $t_{\rm thr} = 4 M_\pi^2$. The spectral functions on the two-pion 
cut can be obtained from the elastic unitarity conditions, which in the $N/D$ 
representation take the form \cite{Hohler:1974eq,Frazer:1960zza,Frazer:1960zzb}
\be
\text{Im} \, G_E^V(t')[\pi\pi] &=& \frac{k_{\rm cm}^3}{m_N \sqrt{t'}} \, J_+^1(t') \, |F_{\pi}(t')|^2 ,  
\label{Eq:ImGEV2}
\\ 
\text{Im} \, G_M^V(t')[\pi\pi] &=& \frac{k_{\rm cm}^3}{\sqrt{2t'}}    \, J_-^1(t') \, |F_{\pi}(t')|^2 ,
\label{Eq:ImGMV2}
\ee
where $k_{\rm cm} = \sqrt{t'/4 - M_\pi^2}$ is the center-of-mass momentum of the $\pi\pi$ 
system in the $t$-channel. Here $J_\pm^1(t') \equiv f_\pm^1(t')/F_\pi(t')$ are the ratios of
the $\pi\pi \rightarrow N\bar{N}$ partial-wave amplitudes and the timelike pion FF,
which are real for $t' > 4 M_\pi^2$ and free of $\pi\pi$ rescattering effects.
These functions can be computed in $\chi$EFT with good 
convergence \cite{Alarcon:2017ivh,Alarcon:2017lhg}. $|F_\pi(t')|^2$ is
the squared modulus of the timelike pion FF, which contains the $\pi\pi$ rescattering
effects and the $\rho$ meson resonance. This function is measured in $e^+e^- \rightarrow \pi^+\pi^-$
exclusive annhihilation experiments with high precision and can be taken from 
a parametrization of the data; see Ref.~\cite{Druzhinin:2011qd} for a review. 
Because the $\pi\pi$ state practically exhausts the $e^+e^-$
annihilation cross section at $t' \lesssim 1$ GeV$^2$, the elastic unitarity relations 
Eqs.~(\ref{Eq:ImGEV2}) and (\ref{Eq:ImGMV2}) are assumed to be valid up to $t' = 1$ GeV$^2$.

The calculation of the $J_\pm^1$ functions in relativistic $\chi$EFT is described in 
Ref.~\cite{Alarcon:2017lhg}. At LO they are given by the $N$ and $\Delta$ Born terms 
in the $\pi\pi \rightarrow N\bar N$ amplitudes and the Weinberg-Tomozawa term. At NLO corrections
arise at tree-level from an NLO $\pi\pi NN$ contact term in the chiral Lagrangian. 
At N2LO pion loop corrections appear, and the structure becomes considerably more complex. 
In Ref.~\cite{Alarcon:2017lhg} we estimated the N2LO corrections to $J_+^1$ by 
assuming that the full N2LO result has the same structure as the tree-level N2LO result,
in which the dominant contribution is the term proportional to $d_1 + d_2$.
No such estimate was performed for $J_-^1$, since its N2LO corrections arise entirely
from loops. In order to extend the reach of our calculation we now want to estimate $J_+^1$ 
and $J_-^1$ at the same level. This becomes possible with a generalizaton of our 
previous arguments. Inspecting the structure of the N2LO loop corrections in 
the $\pi N \rightarrow \pi N$ amplitude, we find that the dominant $t$-channel correction
can be parametrized as
\beq
A^-[\textrm{N2LO loop}] \, = \, 0, \hspace{2em} B^-[\textrm{N2LO loop}] \, = \, \lambda\, t/f_\pi^2 ,
\eeq
where $A$ and $B$ are the invariant amplitudes \cite{Alarcon:2012kn}.
In this form the N2LO loop result in $J_-^1$ has the same structure as a tree-level correction
arising from contact terms, and the parameter $\lambda$ can be determined in the same way 
as in our previous estimate for $J_+^1$.

In order to extend the isovector spectral integrals to masses $t' > 1$ GeV$^2$ we need to 
parametrize the isovector spectral function beyond the two-pion cut. The $e^+e^-$ exclusive
annihilation data show that the isovector cross section above $t' \sim 1$ GeV$^2$ is overwhelmingly
in the $4\pi$ channel and peaks at $t' \approx 2.3$ GeV$^2$ \cite{Druzhinin:2011qd}. 
(Incidentally, this value coincides with
the squared mass of the $\rho'$ resonance observed in the $\pi\pi$ channel.) It is reasonable to 
assume that the strength distribution in the nucleon spectral function follows a similar pattern.
The simplest way to parametrize the high-mass contribution to the isovector spectral function
is by a single effective pole,
\beq
\text{Im} \, G^{V}_{E,M}(t')[\textrm{high-mass}] \; = \; \pi a^{(1)}_{E,M} \, \delta(t' - M_{1}^2) ,
\label{spectral_isovector_highmass}
\eeq
where we choose $M_1^2 = M_{\rho'}^2 = 2.1\, \textrm{GeV}^2$. The total isovector spectral
function is given by the sum of the $\pi\pi$ cut (calculated in DI$\chi$EFT) and the
high-mass part (parametrized by the effective pole),
\beq
\text{Im} \, G^{V}_{E,M} \; = \; 
\text{Im} \, G^{V}_{E,M}[\pi\pi] \; + \; 
\text{Im} \, G^{V}_{E,M}[\textrm{high-mass}] .
\eeq
We then determine the parameters of the N2LO contributions in $G^{V}_{E,M}[\pi\pi]$
and the strength of the effective pole in $G^{V}_{E,M}[\textrm{high-mass}]$
by imposing the sum rules for the isovector charge and magnetic moment, and for
the electric and magnetic radii (here $t_{\rm thr} = 4 M_\pi^2$):
\begin{align}
 &\frac{1}{\pi}\int_{t_{\rm thr}}^\infty dt' \; \frac{\text{Im} \, G_E^V(t')}{t'} 
= {\textstyle \frac{1}{2}} , 
\label{sumrule1_E} \\
 &\frac{1}{\pi}\int_{t_{\rm thr}}^\infty dt' \; \frac{\text{Im} \, G_M^V(t')}{t'} 
= \textstyle{\frac{1}{2}}(\mu^p - \mu^n) , 
\label{sumrule1_M} \\
 &\frac{6}{\pi}\int_{t_{\rm thr}}^\infty dt' \; \frac{\text{Im} \, G_E^V(t')}{t'^2} 
= \langle r^2 \rangle_E^V  
\equiv {\textstyle \frac{1}{2}} [\langle r^2 \rangle_E^p - \langle r^2 \rangle_E^n ] ,
\label{sumrule2_E} \\
 &\frac{6}{\pi}\int_{t_{\rm thr}}^\infty dt' \; \frac{\text{Im} \, G_M^V(t')}{t'^2} 
=  \langle r^2 \rangle_M^V  
\equiv {\textstyle \frac{1}{2}} [\mu^p \langle r^2 \rangle_M^p - \mu^n \langle r^2 \rangle_M^n ]
\label{sumrule2_M} .
\end{align}
Since the charge and magnetic moment are known precisely, the unknown parameters are
essentially determined in terms of the isovector charge and magnetic radii, which can
be allowed to vary over a reasonable range (see below). This makes our parametrization particularly
convenient for applications where the nucleon radii are regarded as basic parameters
or extracted from data.

In the isoscalar FF the lowest singularity is the 3-pion cut ($t_{\rm thr} = 9 M_\pi^2$).
The strength at $t' < 1$ GeV$^2$ is overwhelmingly concentrated in the $\omega$ resonance,
which we describe by a zero-width pole. At $t' \gtrsim 1$ GeV$^2$ the $K\bar K$ and other
channels open up. The exclusive $e^+e^-$ annihilation data show that the strength at 
$t' \sim 1$ GeV$^2$ is concentrated in the $\phi$ resonance \cite{Druzhinin:2011qd}. 
We therefore parametrize the high-mass isoscalar strength by an effective pole at the $\phi$ mass. 
Altogether, our parametrization of the isoscalar spectral function is
\beq
\text{Im} \, G^S_{E,M}(t') \, = \, 
\pi a^\omega_{E,M} \delta(t' - M_\omega^2) + \pi a^\phi_{E,M} \delta(t' - M_\phi^2) .
\label{spectral_isoscalar}
\eeq
The strength of the $\omega$ and high-mass ($\phi$) poles are fixed by imposing the sum rules
for the isoscalar charges and radii, i.e., the analog of Eqs.~(\ref{sumrule1_E})--(\ref{sumrule2_M})
with $V \rightarrow S$ and $(p - n) \rightarrow (p + n)$.

%
%
\begin{table}
\begin{center}
\begin{tabular}{cc}                             
\hline                                                        
$a_E^{(1)}$  & $(-0.853, -0.58 )$ \\  
$a_M^{(1)}$  & $(-2.601, -1.194)$   \\        
\hline                                                        
$a_E^\omega$ & $(0.722, 0.840)$   \\  
$a_M^\omega$ & $(0.613, 0.898 )$   \\        
$a_E^\phi$   & $(-0.905 , -0.705 )$ \\
$a_M^\phi$   & $(-1.064, -0.581)$  \\
\hline
\end{tabular}
\caption{Parameters of the effective poles describing the high-mass isovector
spectral function, Eq.~(\ref{spectral_isovector_highmass}), and the isoscalar
spectral function, Eq.~(\ref{spectral_isoscalar}), as determined by the sum
rule Eqs.~(\ref{sumrule1_E})--(\ref{sumrule2_M}) and the corresponding isoscalar sum rule.
\label{table:poles}}
\end{center}
\end{table}
In fixing the isovector and isoscalar spectral function parameters through the sum rules 
Eqs.~(\ref{sumrule1_E})--(\ref{sumrule2_M}) and their isoscalar analog, we use the
Particle Data Group (PDG) values of the proton and neutron charge radii \cite{Patrignani:2016xqp},
together with a recent dispersive calculation of the isovector charge radius \cite{Hoferichter:2016duk}. 
For the proton and neutron magnetic radii we use the results of 
Refs.~\cite{Lorenz:2012tm, Epstein:2014zua}, which are compatible with the PDG values in 
the neutron case. The empirical variation of the radii generates a range of the parameters,
which then produces the uncertainty bands in our predictions. The resulting parameters
are summarized in Table~\ref{table:poles}. The uncertainty induced by the empirical 
pion timelike FF in the isovector calculation using Eqs.~(\ref{Eq:ImGEV2}) and 
(\ref{Eq:ImGMV2}) is small and can be neglected.

\begin{figure}[t]
\begin{center}
\includegraphics[width=.4\textwidth]{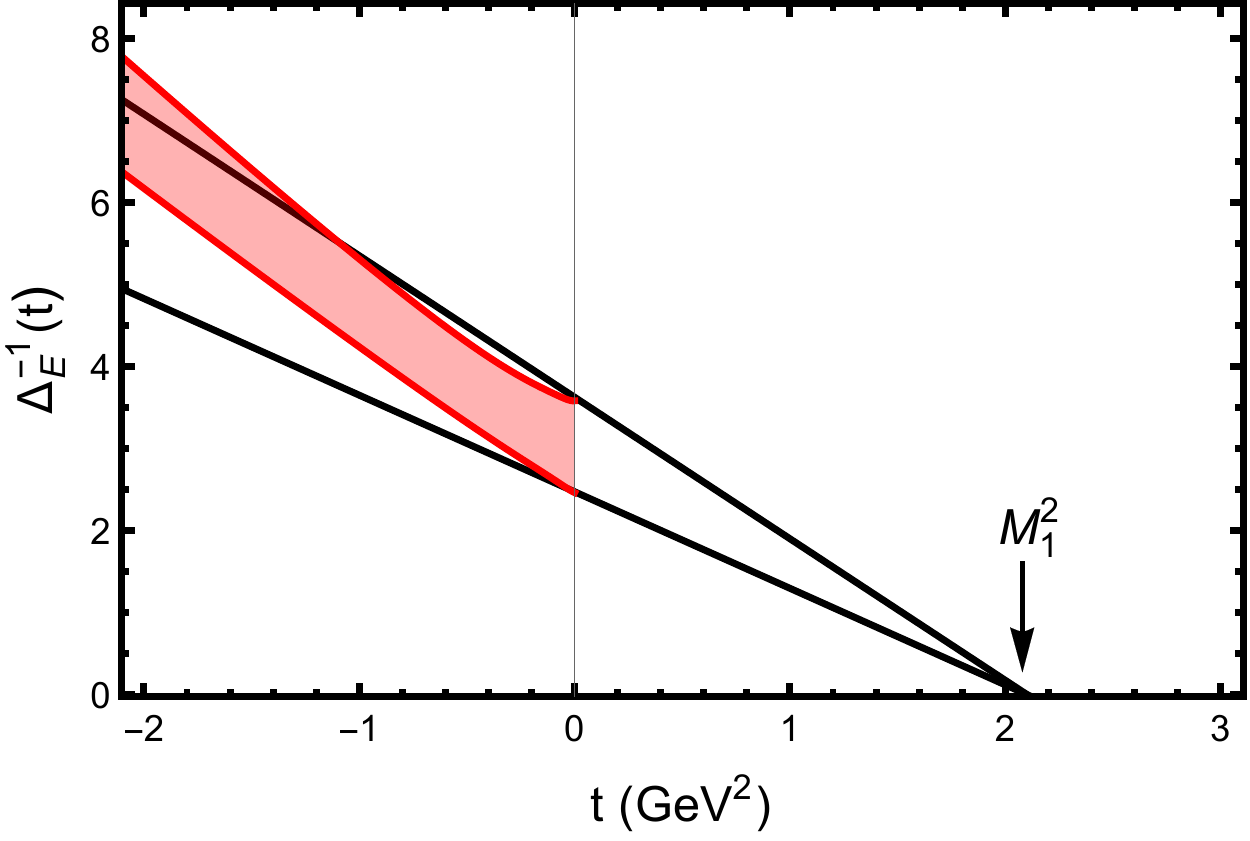}
\end{center}
\vspace{-2ex}
\caption{Red band: Inverse difference $1/\Delta_E(t)$, Eq.~(\ref{diff}). 
Black lines: Inverse of the dispersive integral with the
single-pole parametrization Eq.~(\ref{spectral_isovector_highmass}).}
\label{fig:diff}
\end{figure}

In the present calculation we parametrize the high-mass states in the spectral functions 
by a single effective pole, whose strength can be fixed by the sum rules. The approximation 
is justified as long as we restrict ourselves to the spacelike FFs at moderate momentum 
transfers $|t| \sim$ 1~GeV$^2$. We can demonstrate this explicitly for the isovector FF, 
using a techique described in Ref.~\cite{Hohler:1974eq}. We take the difference of 
the empirical spacelike FF and the finite dispersive integral over the $\pi\pi$ cut
up to $t_{\rm max} = 1\, \textrm{GeV}^2$,
\beq
\Delta_E (t) \, \equiv \, G_E^V (t)[\text{emp}] 
\, - \, \frac{1}{\pi} \int_{t_{\rm thr}}^{t_{\rm max}} dt' \; 
\frac{\text{Im} \, G_E^V(t')[\pi\pi]}{t' - t - i 0} .
\label{diff}
\eeq
This quantity represents the high-mass part of the dispersive integral, which is to be
approximated by the dispersive integral with the effective pole, $a_E^{(1)}/(t - M_1^2)$. 
Plotting $1/\Delta_E(t)$ at $t < 0$ (see Fig.~\ref{fig:diff})
one sees that the dependence on $t$ is approximately linear, and that the single-pole form 
provides an adequate description up to $|t| < 2\, \textrm{GeV}^2$. Note that this is
achieved with the pole parameters fixed by the sum rules Eqs.~(\ref{sumrule1_E})--(\ref{sumrule2_M}),
and that we do not perform a fit of the spacelike FF data in Fig.~\ref{fig:diff}.
%
%

The nucleon FFs obey superconvergence relations
\beq
\int_{t_{\rm thr}}^\infty dt' \; \text{Im} \, G^{V, S}_i(t') \; = \; 0
\hspace{2em} (i = E, M) ,
\label{superconvergence}
\eeq
which guarantee the absence of powers $t^{-1}$ in the asymptotic behavior for
$|t| \rightarrow \infty$. In the present calculation we focus on the FFs at limited spacelike 
momenta $|t| \lesssim 1\, \textrm{GeV}^2$ and are not concerned with the asymptotic behavior.
The relation Eq.~(\ref{superconvergence}) could easily be implemented in our approach 
by parametrizing the high-mass spectral density in a more flexible form; however, 
this would require fitting the spacelike FF data in order to determine the parameters,
which is not our intention here.
\section{Results}
%
%
\begin{figure}[t]
\begin{center}
\includegraphics[width=.43\textwidth]{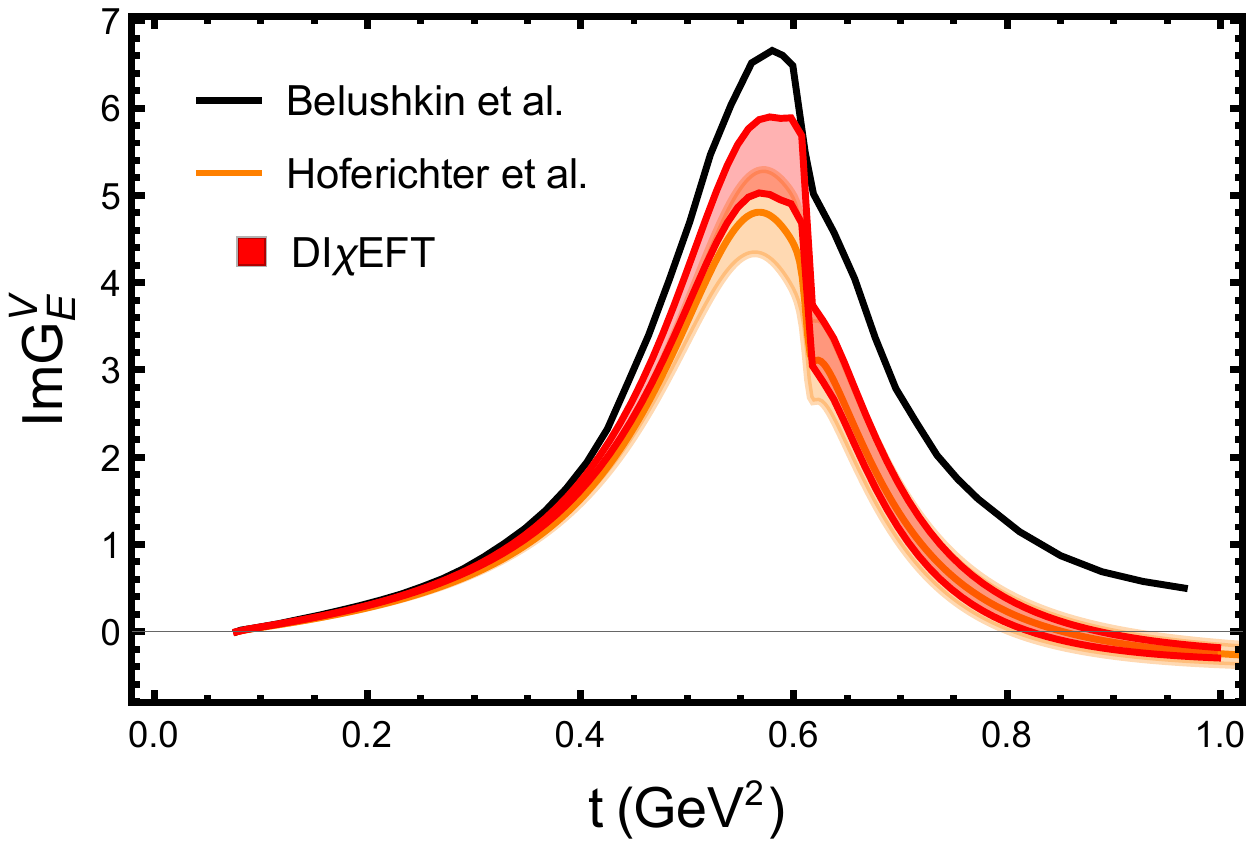} \\
\includegraphics[width=.43\textwidth]{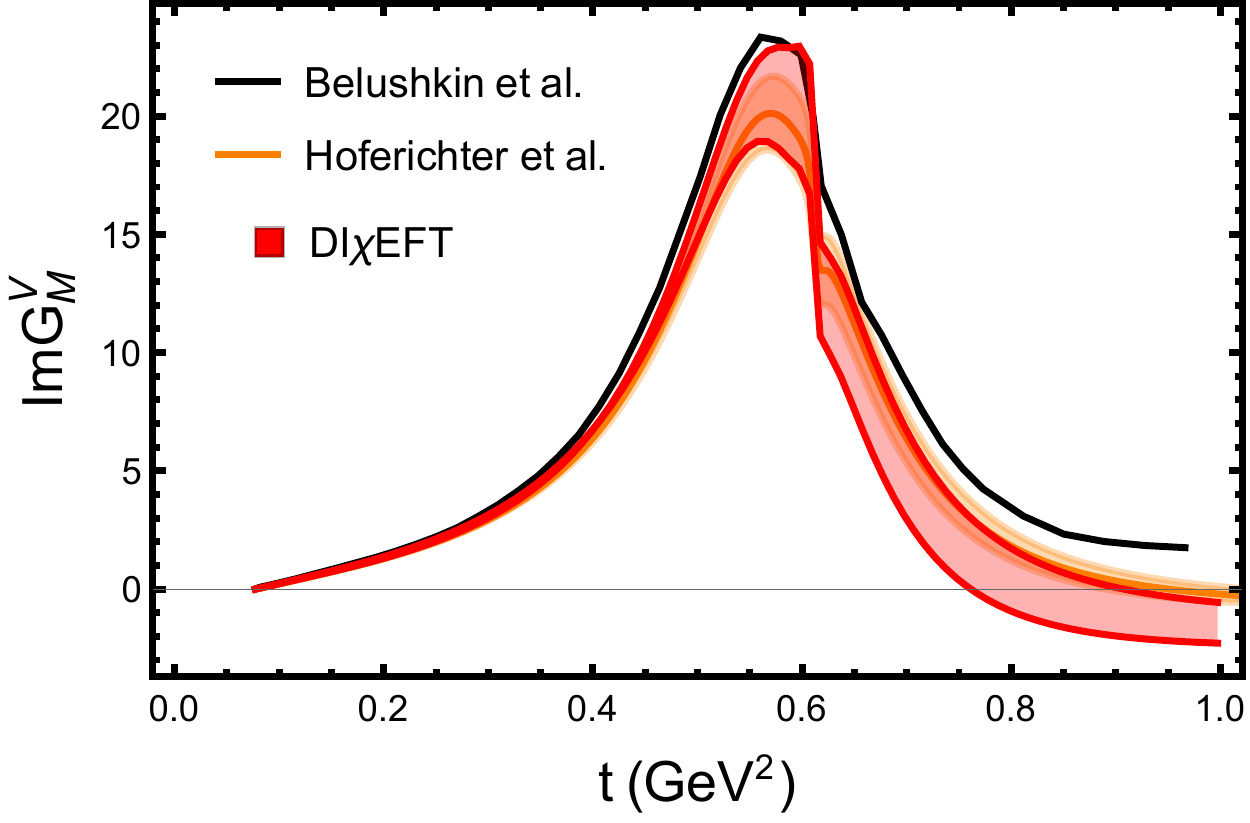}
\vspace{-2ex}
\caption{Red bands: Isovector spectral functions on the two-pion cut calculated in our
approach and their theoretical uncertainty. Orange bands: Spectral functions obtained
in Roy-Steiner analysis of Ref.~\cite{Hoferichter:2016duk}. Black line:
Spectral functions of Ref.~\cite{Belushkin:2005ds}.
\label{fig:spectral}}
\end{center}
\end{figure}
The spectral functions are the primary quantities calculated in our approach. 
The results for the isovector spectral function on the two-pion cut,
Eqs.~(\ref{Eq:ImGEV2}) and (\ref{Eq:ImGMV2}), are shown
in Fig.~\ref{fig:spectral}. The bands show the total uncertainty of our calculation,
resulting from the uncertainty of the low-energy constants in the $\chi$EFT calculation
and the empirical uncertainty of the nucleon radii used to fix the parameters (see above). 
Compared to Ref.~\cite{Alarcon:2017lhg} the electric
and magnetic spectral functions are now calculated at the same order (LO + NLO + partial N2LO).
Both spectral functions now show a trend to negative values above the $\rho$ peak. 
Our results agree overall very 
well with those obtained in an analysis of $\pi N$ scattering data using Roy-Steiner 
equations \cite{Hoferichter:2016duk}; only in the $\rho$ peak our $\textrm{Im}\, G_E^V$ 
is $\sim$15\% larger. Our uncertainties are comparable to those of the Roy-Steiner analysis.
Also shown in Fig.~\ref{fig:spectral} are the empirical spectral functions 
of Ref.~\cite{Belushkin:2005ds}.

%
%
\begin{figure*}[t]
\begin{center}
\includegraphics[width=.43\textwidth]{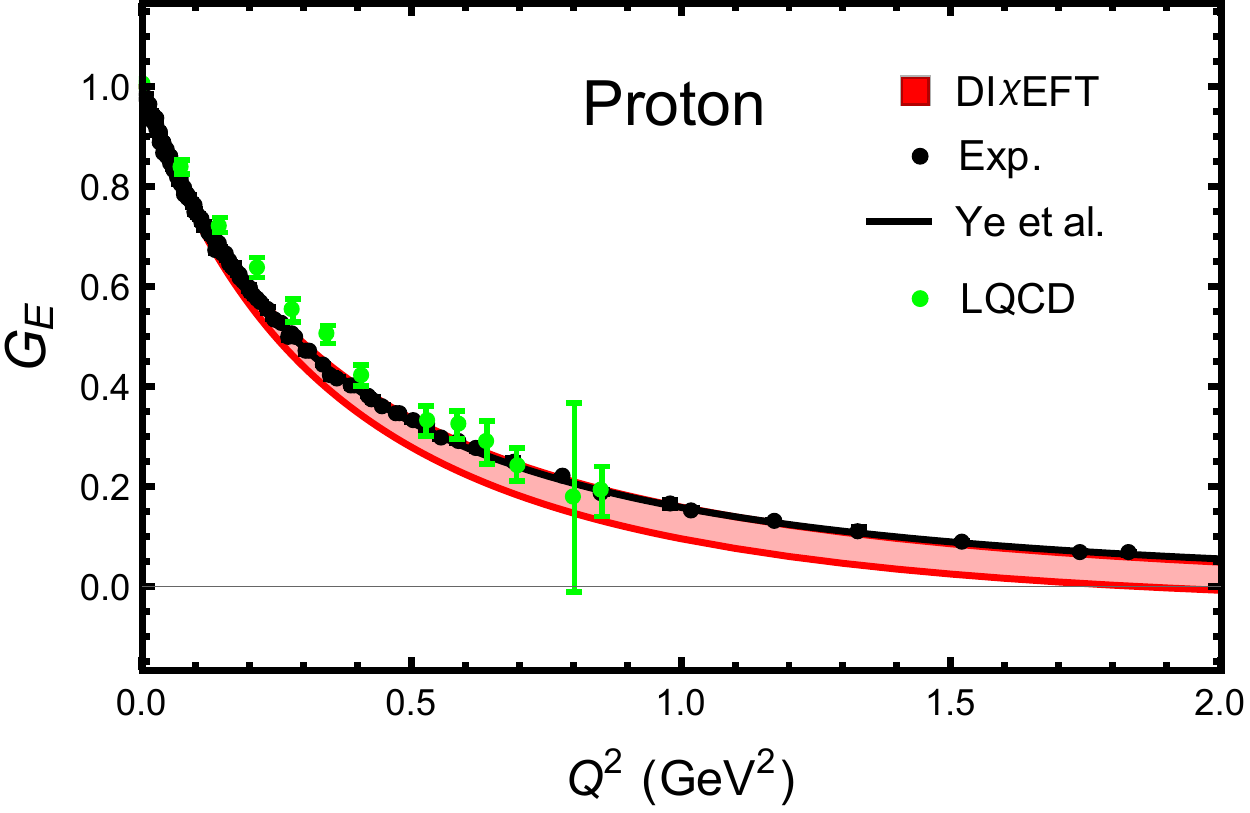}
\includegraphics[width=.43\textwidth]{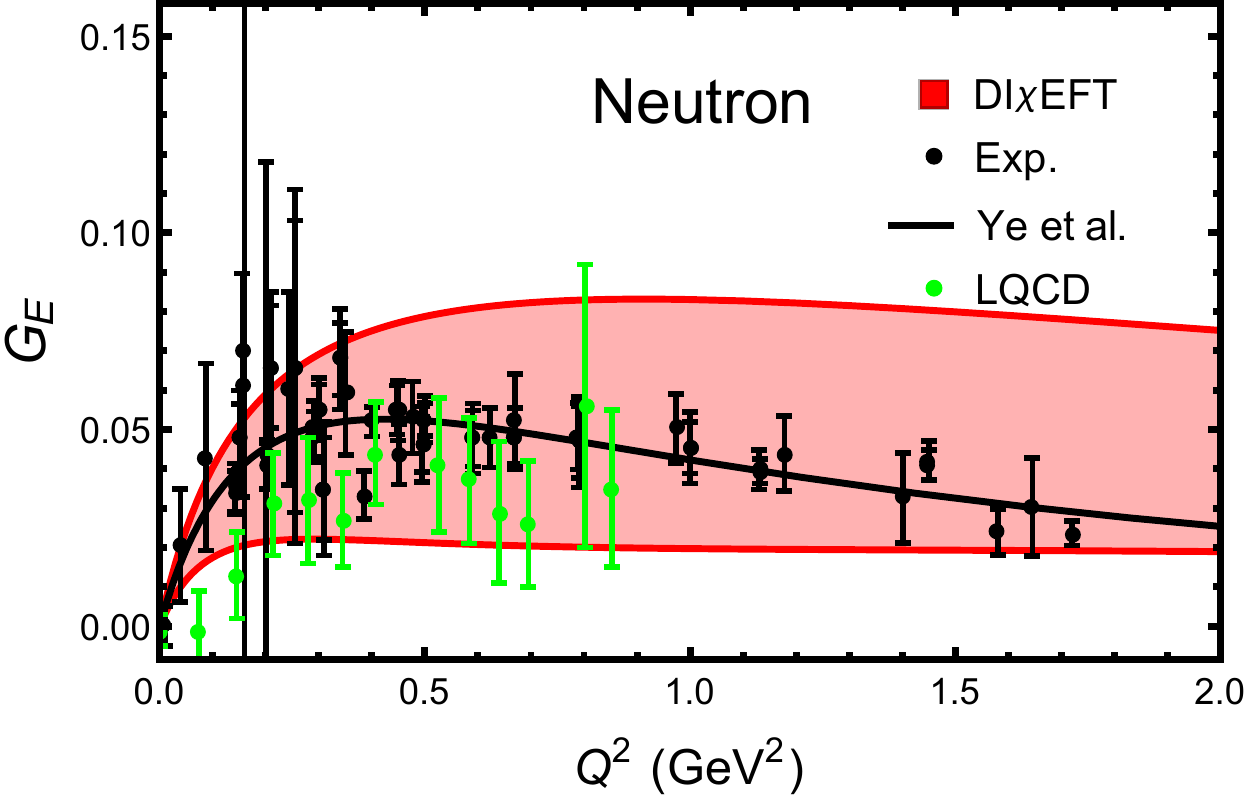} \\
\includegraphics[width=.43\textwidth]{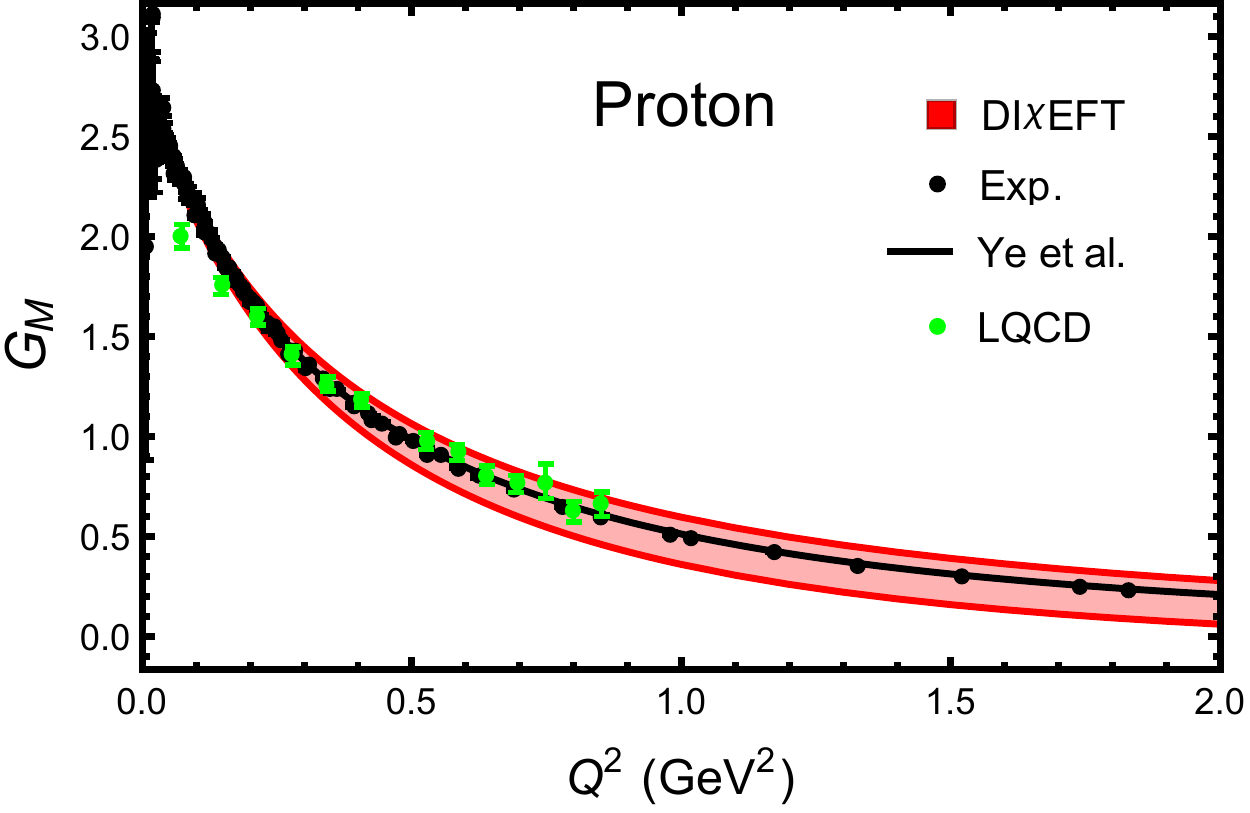}
\includegraphics[width=.43\textwidth]{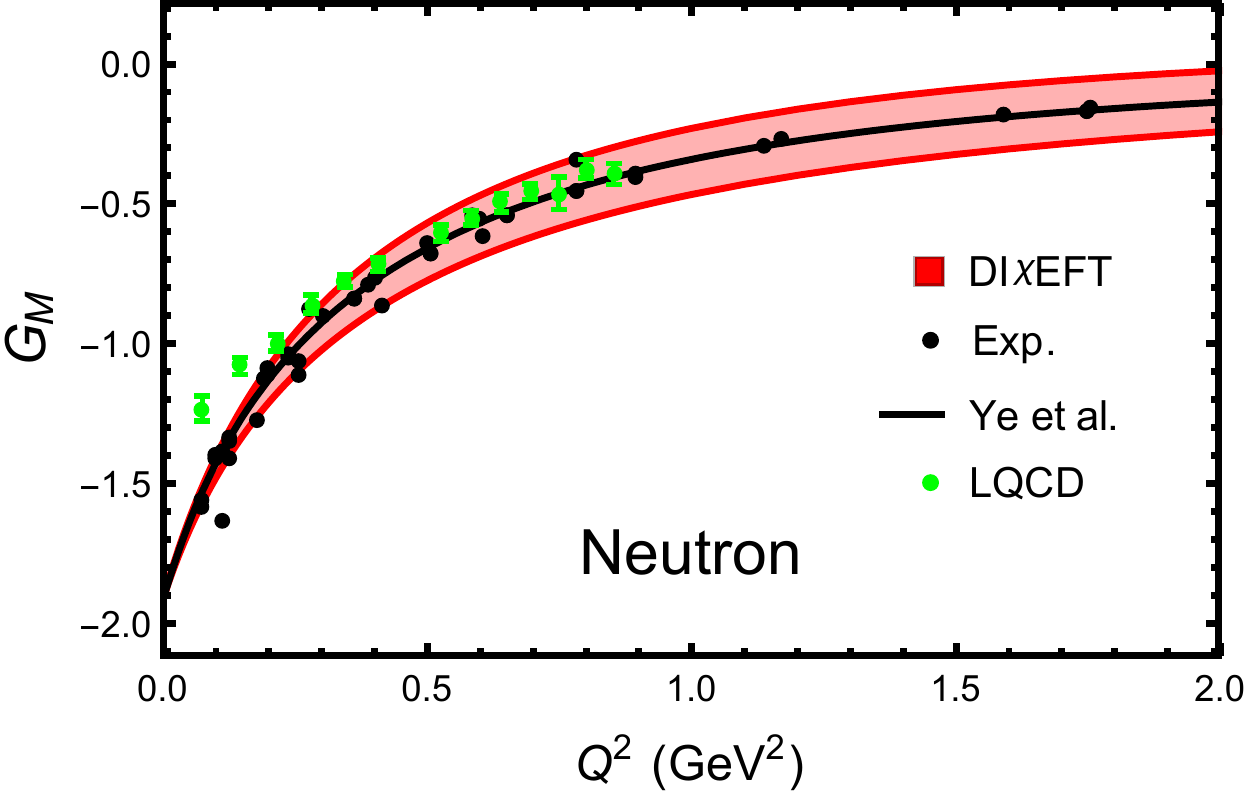}
\caption{Red bands: Proton and neutron EM FFs calculated in our approach and their theoretical
uncertainties. Solid black: Empirical FF parametrization of Ref.~\cite{Ye:2017gyb}.
Black dots: Data of the MAMI A1 experiment \cite{Bernauer:2010wm,Bernauer:2013tpr}.
Green dots: Lattice QCD results from Ref.~\cite{Alexandrou:2017ypw}.}
\label{fig:form}
\end{center}
\end{figure*}
%
%
%
%
\begin{table*}
\begin{center}
\vspace{-2ex}
\begin{tabular}{l|cccc}
\hline
Moment &$G_E^p$ & $G_E^n$ & $G_M^p$ &$G_M^n$ \\
\hline
$\langle r^2 \rangle$ (fm$^2$)$^\ast$                 & (0.701, 0.768) & $(-0.079, -0.146)$  & (0.689, 0.765) &  (0.713, 0.813) \\
$\langle r^4 \rangle$ (fm$^4$)                 & (1.473, 1.602)   & $(-0.635, -0.506)$ & (1.676,  1.782)  &  (2.045,  2.042)  \\
$\langle r^6 \rangle$ (fm$^6$)                 & (8.519, 8.962)  & $(-6.110,  -5.667)$  & (11.525,  11.579)   & (15.231,  15.645)   \\
$\langle r^8 \rangle$ ($10^2$ fm$^8$)         & (1.269, 1.296)  & $(-1.159,   -1.131)$  & (1.834,  1.882)   &  (2.597,  2.691)  \\
$\langle r^{10} \rangle$ ($10^3$ fm$^{10}$)    & (3.933, 3.965)  & $(-3.866,  -3.834)$  & (5.707,  5.905)  &  (8.274,  8.581)  \\
$\langle r^{12} \rangle$ ($10^5$ fm$^{12}$)    & (2.041, 2.049)  & $(-2.039,  -2.031)$  & (2.903,  3.004)  &  (4.233,  4.382)  \\
$\langle r^{14} \rangle$ ($10^7$ fm$^{14}$)    & (1.557, 1.561)  & $(-1.559,  -1.556)$  & (2.158,  2.230)   &  (3.150,  3.255)   \\
$\langle r^{16} \rangle$ ($10^9$ fm$^{16}$)    & (1.624, 1.626)  & $(-1.626,  -1.624)$  & (2.191,  2.260)  &  (3.198,  3.299)   \\
$\langle r^{18} \rangle$ ($10^{11}$ fm$^{18}$) & (2.210, 2.212)  & $(-2.212,  -2.210)$  & (2.905,  2.991)  &  (4.241,  4.367)  \\
$\langle r^{20} \rangle$ ($10^{13}$ fm$^{20}$) & (3.796, 3.799)  & $(-3.799,  -3.796)$ & (4.866,  5.006)  &  (7.105,   7.308)  \\
\hline
\end{tabular}
\end{center}
\vspace{-2ex}
\caption{FF moments obtained from the dispersive integral Eq.~(\ref{dispersive_moments})
with the DI$\chi$EFT spectral functions (LO + NLO + partial N2LO). 
\newline
$^\ast$The $\langle r^2 \rangle$ moments are input values (see text).
\label{tab:moments}}
\end{table*}
The spacelike EM FFs calculated with the dispersion integrals Eq.~(\ref{Eq:Dispersive_representation}) 
are shown in Fig.~\ref{fig:form}. The proton and neutron FFs were obtained as 
$G_i^{p, n} = G_i^S \pm G_i^V\; (i = E, M)$. Contrary to Ref.~\cite{Alarcon:2017lhg} we now do not 
perform any subtractions and calculate the dispersive integral without a cutoff in $t'$,
as the high-mass parts of the spectral functions are now parametrized consistently 
through the effective poles. Our results show excellent agreement with the recent
FF parametrization of Ref.~\cite{Ye:2017gyb} for all momentum transfers $Q^2 \lesssim$ 1 GeV$^2$,
and even up $\sim$2 GeV$^2$, which is remarkable in view of our simple parametrization
of the high-mass spectral functions. Note that $G_E^n$ involves substantial cancellations 
between the isovector and isoscalar components, so that its relative uncertainties are
larger than that of the other FFs.

The higher derivatives of the FFs (moments) are needed in the extraction of the 
proton radius from experimental data. In our dispersive approach they are evaluated as
\beq
\frac{\langle r^{2n} \rangle_i}{(2n+1)!} \; = \; 
\frac{1}{\pi} \int_{t_{\rm thr}}^{\infty} dt' \; \frac{\text{Im} \, G_i(t')}{{t'}^{n+1}} 
\hspace{1em} (i = E, M);
\label{dispersive_moments}
\eeq
see Ref.~\cite{Alarcon:2017lhg} for details. The moments obtained with our spectral functions 
are summarized in Table~\ref{tab:moments}. Compared to the results quoted in 
Ref.~\cite{Alarcon:2017lhg} the isovector LO and NLO parts are exactly the same; the only changes 
are the estimated partial N2LO contributions and the added isovector high-mass contribution.
The isoscalar part is the same as in Ref.~\cite{Alarcon:2017lhg}; only the couplings have now
been determined through the charge and radius sum rules. Our new moments have smaller uncertainty 
than those of Ref.~\cite{Alarcon:2017lhg}. They confirm the ``unnatural size'' of the higher
moments (compared to the dipole expectation) observed in Ref.~\cite{Alarcon:2017lhg}.
\section{Discussion}
DI$\chi$EFT enables first-principles dynamical calculations of the isovector two-pion spectral 
functions with controled uncertainties and results in good agreement with empirical amplitude analysis. 
Together with a minimal effective pole parametrization of the high-mass isovector and isoscalar states,
the method provides an accurate dispersive description of the nucleon FFs up to momentum transfers
$|t| \sim$ 1~GeV$^2$ and above. The method is predictive in the sense that the dynamical input
is provided by chiral dynamics and $e^+e^-$ annihilation data, and no fitting of nucleon FFs 
is performed. This represents major progress in the theory of nucleon FFs
at low momentum transfers.

Our results provide a FF parametrization with exact analyticity in $t$ and can be used 
for theoretical or empirical studies in which this property is essential: (a) Determination
of the peripheral charge and magnetization densities in the nucleon \cite{Alarcon:2017asr}; 
(b) extraction of the proton charge radius from $ep$ elastic scattering data; 
(c) calculation of two-photon-exchange corrections in $ep$ elastic scattering.

This material is based upon work supported by the U.S.~Department of Energy, 
Office of Science, Office of Nuclear Physics under contract DE-AC05-06OR23177.
This work was also supported by the Spanish Ministerio de Econom\'ia y Competitividad and
European FEDER funds under Contract No. FPA2016-77313-P.
\newpage
%
%

%
%
%
\end{document}